\documentclass{ws-ijbc_n}
\usepackage{ws-rotating}     
\usepackage{graphicx}
\begin{document}

\catchline{}{}{}{}{} 

\markboth{M.~V.~Tchakui et al.}{Chaotic dynamics of conservative and time dependent piezoelectric mems hamiltonians }

\title{CHAOTIC DYNAMICS OF PIEZOELECTRIC MEMS BASED ON MAXIMAL LYAPUNOV EXPONENT AND SMALLER ALIGNMENT INDEX COMPUTATIONS}
\author{M.~V.~Tchakui\footnote{Author for correspondence} \and P.~Woafo}

\address{Laboratory of Modelling and Simulation in Engineering, Biomimetics\\ and Prototypes,Department of Physics,Faculty of Science,\\ University of Yaounde I, BOX 812\\
Yaounde,  Cameroon\\
muriellevan@yahoo.fr* and pwoafo1@yahoo.fr}  
\author{Ch.~Skokos}
\address{Nonlinear Dynamics and Chaos group, Department of Mathematics\\ and Applied Mathematics, University of Cape Town\\
Rondebosch 7701, South Africa\\
haris.skokos@gmail.com}

\maketitle


\begin{abstract}
We characterize the dynamical states of a piezoelectric microelectromechanical system (MEMS) using several numerical quantifiers including the maximal Lyapunov exponent, the Poincar\'{e} Surface of Section and a chaos detection method called the Smaller Alignment Index (SALI). The analysis makes use of the MEMS Hamiltonian. We start our study by considering the case of a conservative piezoelectric MEMS model and  describe the behavior of some representative phase space orbits  of the system. We show that  the dynamics of the piezoelectric MEMS  becomes considerably more  complex as the natural frequency of the system's mechanical part decreases.This refers to the reduction of the stiffness of the piezoelectric transducer. Then, taking into account the effects of damping and time dependent forces on the piezoelectric MEMS, we derive the corresponding non-autonomous Hamiltonian and investigate its dynamical behavior. We find that the non-conservative system exhibits a rich dynamics, which is strongly influenced by the values of the parameters that govern the piezoelectric MEMS energy gain and loss. Our results provide further evidences of the ability of the SALI to efficiently characterize the chaoticity of dynamical systems.
\end{abstract}

\keywords{Hamiltonian systems; Piezoelectric MEMS; SALI; Lyapunov exponent; Chaos; Regular dynamics.}

\begin{history}
\textit{Preprint accepted for publication in the International Journal of Bifurcation and Chaos, December 6, 2019}
\end{history}
\section{Introduction}
\label{intro}
 In many cases the dynamical behavior of physical systems can be modeled by Hamiltonian systems. Over the years the Hamiltonian formulation has been successfully applied in numerous areas of physics such as statistical mechanics \cite{Brody}, classical physics \cite{Rohrlich}, quantum mechanics \cite{Arminjon} and many other fields \cite{Bountis_springer}. In general, Hamiltonian systems can be divided in two broad categories: conservative and non-conservative systems.  A system is said to be conservative when the value of the corresponding Hamiltonian function (which is usually referred as the system's total energy) remains constant throughout  time. As a typical example of this kind let us mention the well-known H\'{e}non-Heiles system,  which describes, at some approximation, the motion of stars around a galactic center \cite{Henon}. Non-conservative Hamiltonians can describe systems in the presence of external forces depending on time (time dependent Hamiltonian systems) and/or friction forces  (dissipative Hamiltonian systems) provoking the change of the systems total energy.

      In this study, we  focus our attention  on the dynamics of piezoelectric micro-electromechanical systems (MEMSs),
  whose behavior can be described by conservative or non-conservative Hamiltonians depending on the assumptions made for the MEMSs performance. In MEMSs, the piezoelectric effect is used in one of the following ways: applying a mechanical stress to piezoelectric materials produces an electrical charge; or conversely, an applied electrical voltage produces a mechanical strain or motion in a piezoelectric material \cite{Crawley,Schaffner,WENDELL}. The second situation is known as the inverse piezoelectric effect and is the main topic of the present work.  Since its discovery in 1880 \cite{curie1,curie2},  the piezoelectric effect has evolved from a laboratory curiosity to a mature technology. Piezoelectric sensors and actuators are common in sonar systems, proximity sensors, pressure sensors, ink jet printers, speakers, microphones and many other applications \cite{Kuntzman,Philipps,Dakua,Ueberschlag,Risio}.   Investigating the dynamical behavior of piezoelectric MEMSs, as in this work, will assist us to better understand the functioning of devices using piezoelectric actuators.
  
  The study of the dynamical properties of  Hamiltonian systems  constitutes an important research topic  in nonlinear physics, because such systems can exhibit very complex and quite interesting behaviors. Several theoretical and numerical tools have been developed and applied by many researchers in order to investigate the chaotic  dynamics of Hamiltonian systems. Let us briefly present some of them. The numerical construction of the so-called Poincar\'{e} Surface of Section (PSS) has been used to reveal the chaotic properties of mainly non-integrable two degrees of freedom (2dof) Hamiltonian systems, as its extension to higher dimensional models can become problematic (see for example Sect.~1.2 of Ref.~\cite{lieberman92}). The computation of the maximum Lyapunov exponent (mLE) \cite{Benettin1,Benettin2,Skokos_lect_lyap} is the most commonly used method to characterize chaos.
  More recently, several other chaos detection methods have been proposed in the literature, such as the  Fast Lyapunov Indicator (FLI) \cite{Froeschl_lega_2000,Froeschl_lega_2001} and its variants \cite{B_05,B_06}, the Mean Exponential Growth of Nearby Orbits (MEGNO) \cite{CS_00,CGS_03}, the Relative Lyapunov Indicator (RLI) \cite{SEE_00,SESF_04}, as well as the Smaller Alignment Index (SALI) \cite{Skokos_alignment,Skokos_al_prog,Skokos_al_jpa} and its extension the Generalized 
  Alignment Index (GALI) \cite{Skokos_al_physD,SBA_08,MSA_12,Skokos_manos}, to name a few. Review presentations of these, as well as of some other commonly used chaos detection techniques, can be found in \cite{SGL_16}. The SALI proved to be a simple, fast and efficient tool for distinguishing between ordered and chaotic motions, and has already been successfully applied to several models \cite{Bountis_skokos,ABS06,Manos_al_npcs} (see also the review paper of Skokos and Manos \cite{Skokos_manos} and references therein). The performance  of the SALI for dissipative or time dependent systems has also been studied \cite{HSP19,Huang_wu_lect,Huang_wu,Huang_zhou,Manos_al_jpa,Huang_cao}. In these works it has been found that the SALI  behavior is similar to the one shown in the case of conservative systems, and that the index remains an  efficient and accurate tool for detecting chaos in non-conservative systems.

  In the present paper we use the PSS, the mLE and the SALI techniques to investigate the chaotic dynamics of a time dependent piezoelectric MEMS. In \cite{Taffoti} this system was studied in the framework of the Lagrangian formalism, but only its chaotic state was analyzed. Here, a Hamiltonian formulation of the problem is derived taking into account dissipation (friction) and time dependent forces.  Moreover, the analysis of the system's global dynamics is discussed in detail.  The paper is organized as follows: Section \ref{conser_H} deals with the dynamics of the conservative version of the piezoelectric MEMS. Section \ref{nonconser_H} is devoted to the case of the time dependent form of the system, while in Section  \ref{conclu} we summarize our results and present the conclusions of our work.
%
  \section{\label{conser_H}The conservative Hamiltonian piezoelectric MEMS model}
  
   \subsection{\label{Ham_cons}Model and Hamiltonian function}
\begin{figure}[h]
\centering
 \begin{center}
\includegraphics[scale=0.5] {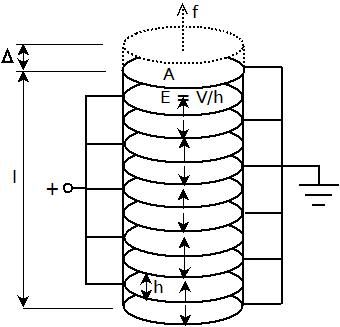} \textbf{(a)}
\includegraphics[scale=0.6] {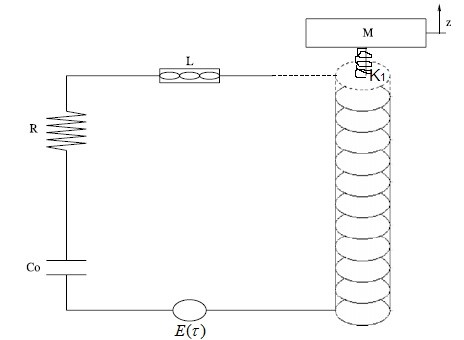}\textbf{(b)}
\end{center}
\caption{(a) A stack of $n$ disks of thickness $h$ and cross section $A$ constituting one piezoelectric actuator. This piezoelectric material is subjected to a total voltage $V$ inducing the electric field $E=V/h$. The total force $f$ resulting from the electric field  produces a total mechanical displacement of the structure $\Delta$. $l$ is the length of transducer. The doubled arrowed segments in the middle of the discs indicate the polarization of the piezosystem. (b) The electrical circuit equipped with the piezoelectric body in oscillation. The piezoelectric transducer is connected to a voltage source $E$, which  varies in time $\tau$, in series with a resistor $R$, an inductor $L$ and a  capacitor $C_0$. The  structure has at its one end spring of stiffness $K_1$ attached to a movable mass $M$ which can be displaced along the $z$ direction.}
\label{model}
\end{figure}

A MEMS is a physical system whose dimensions are of the order of the micrometer. It is made of a mechanical part (flexible or rigid structures) and an electrical part. The model of the  piezoelectric  MEMS considered in this study is presented in Fig.~\ref{model}(a). Following \citet{Taffoti}, we model one piezoelectric MEMS element as a stack of $n$ disks of thickness $h$  and cross section $A$, assuming that all the electrical and mechanical quantities are uniformly distributed in the linear transducer. This model can be found in technological devices where the inverse piezoelectric effect is brought into place. When the piezoelectric element is subjected to a voltage $V$ it exhibits a displacement $\Delta$ which is proportional to the input signal. For this study, the piezoelectric transducer is connected to a voltage source 
\begin{eqnarray}
E(\tau ) = {E_{e0}}\cos {\omega _0}\tau,
\label{voltage}
\end{eqnarray}
in series with a resistor $R$, an inductor $L$ and a nonlinear capacitor [see Fig.~\ref{model}b] whose charge-voltage characteristic is given by: 
\begin{eqnarray*}
V_{C0} = q/C_{0l} + \beta_{e0}q^3,
\nonumber
\end{eqnarray*}
where $C_{0l}$ and $\beta_{e0}$ are respectively the linear value of the capacitor $C_0$ and the nonlinear coefficient. The piezo structure is also equipped at one end of a spring with nonlinear stiffness $K_1$ as presented in Fig.~\ref{model}(b). 

The total extension of the piezosystem $\Delta$ can be expressed as:
\begin{eqnarray*}
\Delta = bz,
\end{eqnarray*}
where $b$ is a coefficient relating the end displacement of the transducer to the global coordinate system $z$. The total dissipation of the system which is the sum of the mechanical loss resulting from internal damping and the electrical loss due to the Joule's effect in the resistor $R$ is given by the following function: 
 \begin{equation}
 \Lambda  = \frac{1}{2}{\lambda _{m0}}{\dot z^2} + \frac{1}{2}R{\dot q^2},
 \label{dissipation}
 \end{equation}
where $\lambda _{m0}$  is damping coefficient. The used piezoelectric actuator which is of \textit{lead zirconate-titanate} (PZT) type commonly achieve a relative displacement of up to $0.2{\raise0.5ex\hbox{$\scriptstyle 0$}\kern-0.1em/\kern-0.15em\lower0.25ex\hbox{$\scriptstyle 0$}}$. The values of the physical parameters of the piezoelectric transducer are presented in  Table \ref{tab:value}.

\begin{table}[h]
\tbl{\label{tab:value} Parameter values of the piezoelectric transducer.}
{\begin{tabular}{c c c}\\[-2pt]
\toprule
Property & Symbol & Value and Unit \\
\hline\\[-2pt]
Number of disks & $n$ & 1530\\
Thickness of one disk & $h$ & $3 ~\mu m$\\
Total length of transducer & $l=nh$ & $4.59 ~mm$\\
Diameter of the transducer & $D$ & $4 ~mm$\\
Piezoelectric constant & $d_{33}$ & $300 \times {10^{ - 12}} ~C/N$\\
Stiffness for small stretching & $K_0$ & $6.67 ~N/m$\\
Mechanical nonlinear coefficient & & \\
related to the stiffness & $K_1$ & $8.39 ~N/{m^3}$\\
Density mass & $\rho$ & $7600 ~kg/{m^3}$\\
Electromechanical coupling factor & $k$ & $0.4$\\
Young modulus & ${\nu _E} $ & $50 \times {10^9}~Pa$\\
Dielectric constant under & & \\
constant stress & $\varepsilon ^T$ & $1.593 \times {10^{ - 8}} ~F/m$\\
Capacitance of the transducer & & \\ with no external load & $C $ & $100 ~\mu F$\\
Electrical nonlinear coefficient & & \\  related to the capacitor & $\beta_{e0} $ & $150 ~V C$\\
Viscous damping coefficient & $\lambda_{m0} $ & $0.0093 ~{N s}/m$\\
 Voltage source amplitude & $E_{e0}$ & $109.5 ~V$\\
 Resistance & $R$ & $0.17 ~\Omega$\\
Inductance & $L$ & $1 ~H$\\
Linear value of capacitor $C_0$ & $C_{0l}$ & $1 ~F$\\
\botrule
\end{tabular}}
\end{table}

The Lagrangian $\Gamma$  of the piezoelectric MEMS is \cite{Taffoti}:
\begin{eqnarray}
\Gamma  = \frac{1}{2}M\dot z^2  + \frac{1}{2}L\dot q^2  - \frac{1}{2}\left( {K_0  + \frac{{K_a b^2 }}
{{1 - k^2 }}} \right)z^2  - \frac{1}{4}K_1 z^4 \nonumber\\
 - \frac{1}{2}\left( {\frac{1}{{C_{0l} }} + \frac{1}
{{C(1 - k^2 )}}} \right)q^2  - \frac{1}{4}\beta _{e0} q^4  + \frac{{nd_{33} K_a b}}
{{C(1 - k^2 )}}qz,
\label{eq1}
\end{eqnarray}
where $q$  and $z$ are variables respectively related to the electrical charge and the mechanical displacement which vary according to time $\tau$ .  $M$ is the mass of the structure, $K_a={A\nu_E}/l$  is the stiffness with short circuited electrodes.

We conduct some mathematical transformations of Eq.~(\ref{eq1}), as shown in Appendix A, and get the following Hamiltonian function, which describes the system's dynamics in  dimensionless variables:

\begin{eqnarray}
H\left( {{p_q},{p_z},q,z} \right) = \frac{{{\beta _1}}}{{300}}p_q^2 + \frac{{{\gamma _2}{\beta _1}}}{{300{\gamma _1}}}p_z^2 + \frac{{75}}{{{\beta _1}}}{q^2} + \frac{{75}}{2}{q^4}\nonumber\\
 + \frac{{75{\gamma _1}\omega _2^2}}{{{\gamma _2}{\beta _1}}}{z^2} + \frac{{75{\gamma _1}{\beta _2}}}{{2{\gamma _2}{\beta _1}}}{z^4} - \frac{{150{\gamma _1}}}{{{\beta _1}}}qz.
\label{eq3}
\end{eqnarray}
Here $q$,  $z$ are the  generalized coordinates and $p_q$, $p_z$  the  generalized momenta of respectively the electrical and mechanical parts of the system. In addition, $\gamma_1$ and  $\gamma_2$ are the electromechanical coupling coefficients, $\beta_1$ and $\beta_2$ are the nonlinearity coefficients, while $\omega_2$ is the natural frequency of the mechanical part.  The expressions of all these quantities are:
\begin{eqnarray*}
 {\gamma _1} = \frac{{n{d_{33}}{K_a}b}}{{LC{\omega_e}^2\left( {1 - {k^2}} \right)}},~~
 {\gamma _2} = \frac{{n{d_{33}}{K_a}b}}{{MC{\omega_e}^2\left( {1 - {k^2}} \right)}},~~ {\beta _1} = \frac{{{\beta _{e0}}}}{L{\omega_e}^2},\\ {\beta _2} = \frac{K_1}{M{\omega_e}^2},~~ \omega _2^2 = \frac{1}{M{\omega_e}^2}\left( {{K_0} + \frac{{{K_a}{b^2}}}{{1 - {k^2}}}} \right),\\  \textrm{with}~~\omega _e^2 = \frac{1}{L}\left( {\frac{1}{{{C_{0l}}}} + \frac{1}{{C(1 - {k^2})}}} \right).
\end{eqnarray*}

We note that Hamiltonian (\ref{eq3}) is a 2dof autonomous  system (i.~e.~it does not explicitly depend on the dimensionless time $t$),  governed by the following equations of motion:
\begin{eqnarray}
 \dot q &=& \frac{{\partial H}}{{\partial {p_q}}} = \frac{{{\beta _1}}}{{150}}{p_q}{\rm{ }}   \nonumber\\
\dot z &=& \frac{{\partial H}}{{\partial {p_z}}} = \frac{{{\gamma _2}{\beta _1}}}{{150{\gamma _1}}}{p_z}  \nonumber\\
{{\dot p}_z} &=&  - \frac{{\partial H}}{{\partial z}} =  - \frac{{150{\gamma _1}\omega _2^2}}{{{\gamma _2}{\beta _1}}}z - \frac{{150{\gamma _1}{\beta _2}}}{{{\gamma _2}{\beta _1}}}{z^3} + \frac{{150{\gamma _1}}}{{{\beta _1}}}q \nonumber\\
{{\dot p}_z} &=&  - \frac{{\partial H}}{{\partial z}} =  - \frac{{150{\gamma _1}\omega _2^2}}{{{\gamma _2}{\beta _1}}}z - \frac{{150{\gamma _1}{\beta _2}}}{{{\gamma _2}{\beta _1}}}{z^3} + \frac{{150{\gamma _1}}}{{{\beta _1}}}q,
\label{eq4}
\end{eqnarray}
where dot $(\,\,\dot{} \,)$ denotes the time derivative.

In order to determine the regular or chaotic nature of orbits by the computation of the mLE and/or the SALI, we need to follow the time evolution of small deviations from the considered orbits. In other words, we need to consider in time a deviation vector $\vec{w}(t)$ having as coordinates the small variations $\delta q$, $\delta z$, $\delta p_q$, $\delta p_z$ of variables $q$, $z$, $p_q$, $p_z$ respectively, i.~e.~$\vec{w}(t) = \left( \delta q (t), \delta z (t), \delta p_q (t), \delta p_z (t) \right)$. The evolution of these deviations is governed by the so-called variational equations of the system (see for example \cite{Skokos_lect_lyap}). The variational equations of the Hamiltonian (\ref{eq3})  are as follows:
\begin{eqnarray}
  \dot{\delta q} &=& \frac{{{\beta _1}}}{{150}}\delta {p_q}  \nonumber \\
  \dot{\delta z} &=& \frac{{{\gamma _2}{\beta _1}}}{{150{\gamma _1}}}\delta {p_z}  \nonumber \\
  \dot{\delta p_q}  &=&  - \left( {\frac{{150}}{{{\beta _1}}} + 450{q^2}} \right)\delta q + \frac{{150{\gamma _1}}}{{{\beta _1}}}\delta z \nonumber \\
  \dot{\delta  p_z} & =& - \left( {\frac{{150{\gamma _1}\omega _2^2}}{{{\gamma _2}{\beta _1}}} + \frac{{450{\gamma _1}{\beta _2}}}{{{\gamma _2}{\beta _1}}}{z^2}} \right)\delta z + \frac{{150{\gamma _1}}}{{{\beta _1}}}\delta q.
\label{eq5}
 \end{eqnarray}
We note that the variational equations (\ref{eq5}) cannot be solved independently from the equations of motion (\ref{eq4}) as they explicitly depend on variables $q$ and $z$. Thus,  equations (\ref{eq4})  and (\ref{eq5}) have to be solved simultaneously and be treated as one, large set of differential equations. In our study, we numerically solve this set by using the fourth order Runge-Kutta method with a time step $10^{-3}$.

\subsection{\label{sec:level3}Dynamics}
Based on the analysis presented in ~\cite{Taffoti}, we set the values  of the parameters of  Hamiltonian (\ref{eq3}) to
\begin{eqnarray*}
\beta_1  = 14.25,~ \beta_2  = 13.91,~ \gamma _1  = 0.21,~ \gamma _2  = 3.64~ \textrm{and}~ \omega _2^2  = 3.75
\end{eqnarray*}
for our investigation. 
In all our simulations the absolute value of the relative energy error
\begin{eqnarray*}
E_r=\left| \left[ H(t)-H(0)\right]/ H(0)\right|,
\end{eqnarray*}
where $H(t)$ and $H(0)$ are the values of  Hamiltonian (\ref{eq3}) at times $t=0$ and $t>0$ respectively, remains always below $10^{-10}$. This clearly indicates the very good accuracy of our computations.

\begin{figure*}[h]
\centering
\includegraphics[width=0.38\textwidth,height=0.37\textwidth] {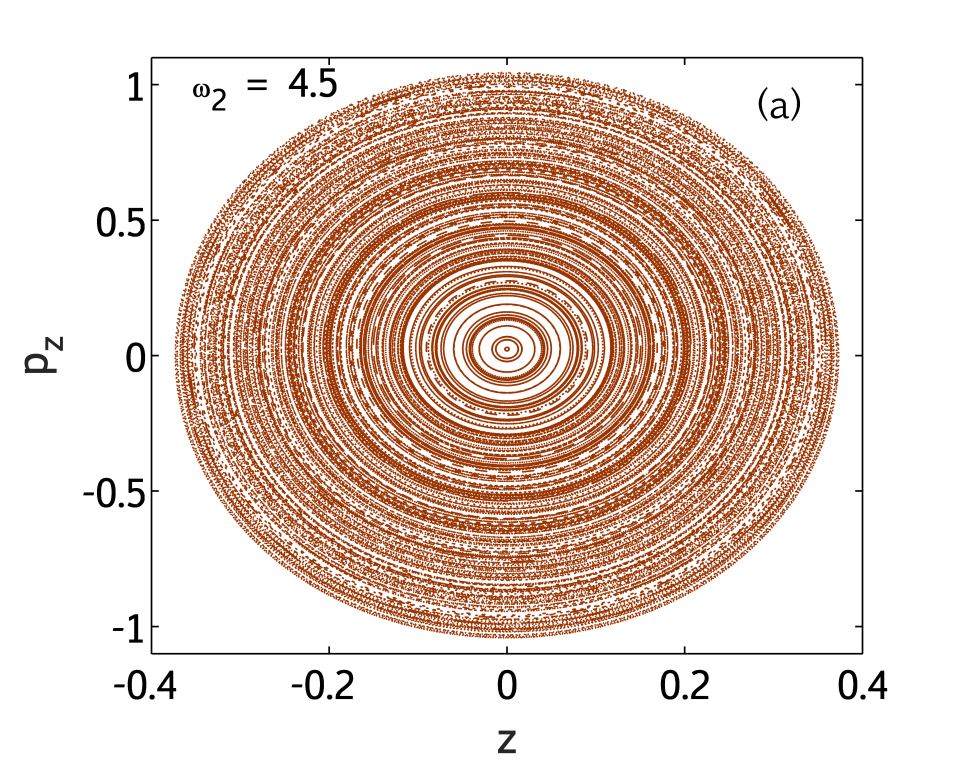} 
\includegraphics[width=0.38\textwidth,height=0.37\textwidth] {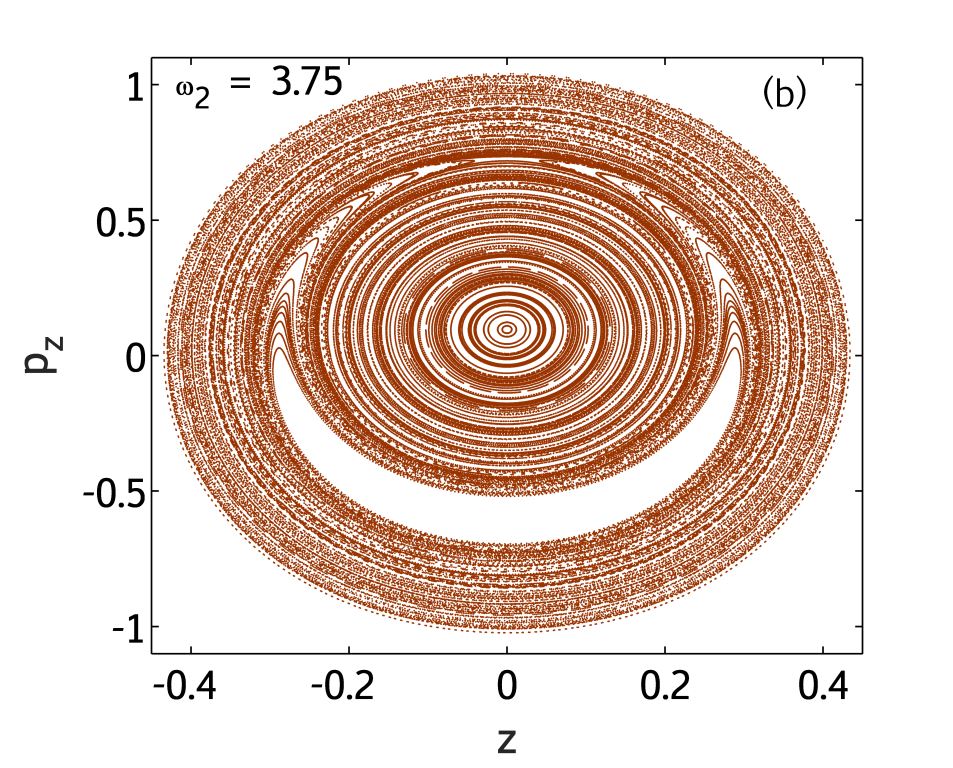} 
\includegraphics[width=0.38\textwidth,height=0.37\textwidth] {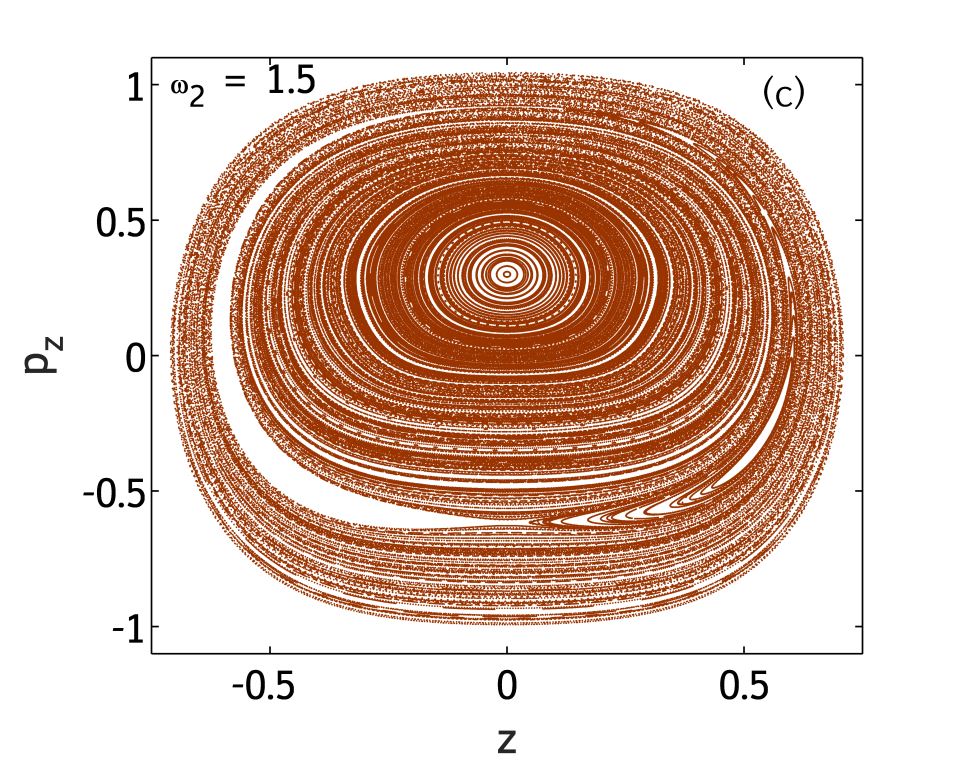} 
\includegraphics[width=0.38\textwidth,height=0.37\textwidth] {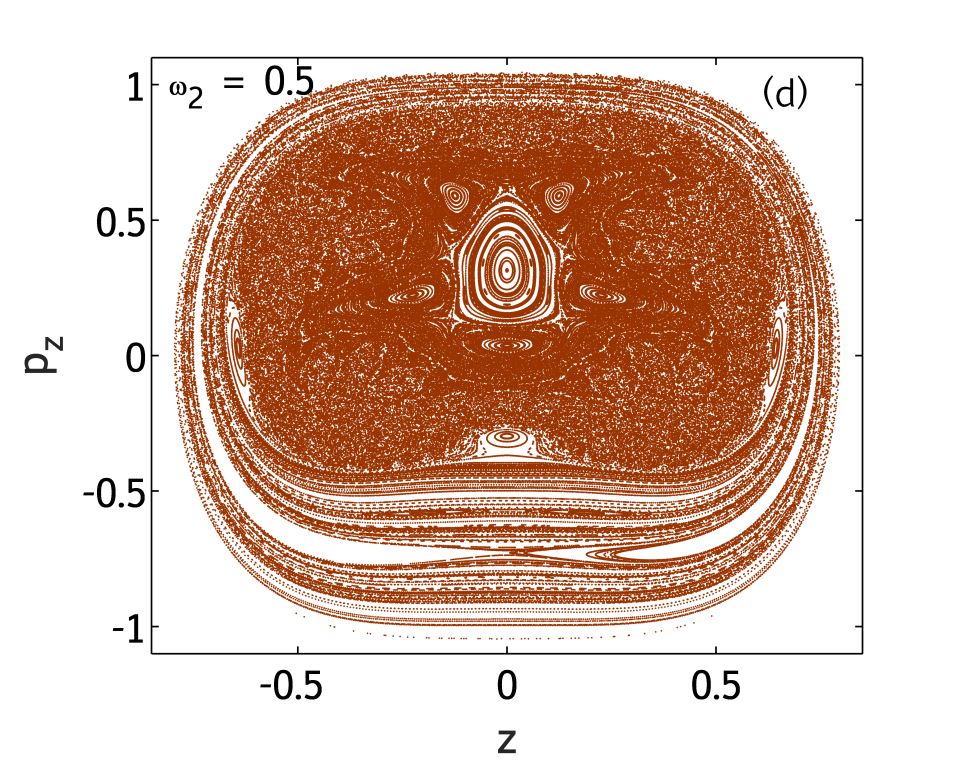} 
 \begin{center}
\includegraphics[width=0.43\textwidth,height=0.42\textwidth] {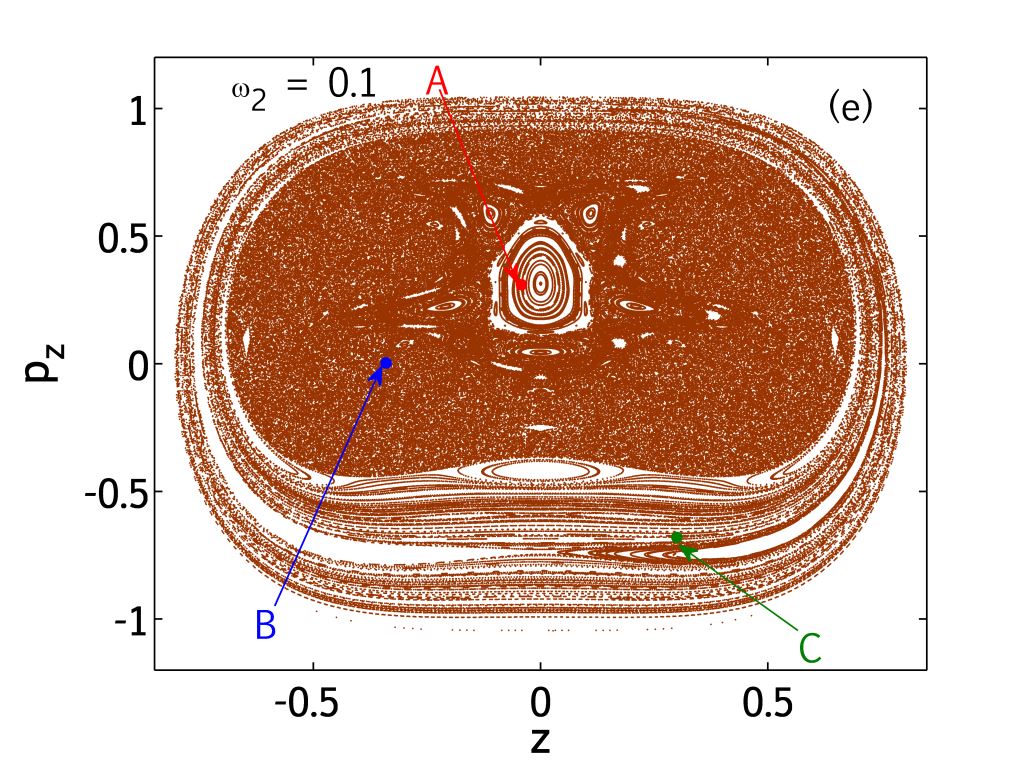} 
\end{center}
\caption{The PSS $(z,p_z)$ of the Hamiltonian model (\ref{eq3}) for $q = 0$, $p_q > 0$ and various values of the mechanical natural frequency: (a) $\omega_2 = 4.5$, (b) $\omega_2 = 3.75$, (c) $\omega_2 = 1.5$, (d) $\omega_2 = 0.5$ and (e) $\omega_2 = 0.1$. In (e), the initial conditions of a regular orbit (A), a chaotic one (B) and a regular orbit (C)  are denoted respectively  by red, blue and green color  dots and indicated by arrows.}
\label{energy}
\end{figure*}

In Fig.~\ref{energy} we plot the system's PSS for different values of its mechanical natural frequency. The PSS is obtained by plotting the $z$ and $p_z$ coordinates of the intersections of several orbits with the phase subspace defined by $q=0$ and $p_q>0$. A grid of $50 \times 50$ equally spaced initial conditions in the $(z,p_z)$ plane is considered in each panel.

For $\omega_2 = 4.5$ [Fig.~\ref{energy}(a)],  $\omega_2 = 3.75$ [Fig.~\ref{energy}(b)] and $\omega_2 = 1.5$ [Fig.~\ref{energy}(c)] we see that the phase space is mainly occupied by invariant curves, which correspond   to the intersections of two-dimensional tori of quasiperiodic motion with the PSS, indicating that the dynamics is predominately  characterized by regular motions.

For smaller values of the mechanical natural frequency ($\omega_2 = 0.5$ [Fig.~\ref{energy}(d)] and $\omega_2 = 0.1$) [Fig.~\ref{energy}(e)], which correspond to the stiffness decrease of the considered piezoelectric transducer, more complicated pictures are seen: regions of regular motion, corresponding to what looks to be smooth curves, coexist with scattered points belonging to chaotic orbits.

Let us consider three representative orbits A, B and C of that system having $\omega_2 = 0.1$ [Fig.~\ref{energy}(e)] and a total energy $H=0.9$ with the following initial conditions
\begin{eqnarray*}
 \textrm{Orbit A (regular)}~&:&~ q = 0;~ z = -0.042;~ p_z = 0.31,\\
 \textrm{Orbit B (chaotic)}~&:&~ q = 0;~ z = -0.34;~ p_z = 0.0032,\\
 \textrm{Orbit C (regular)}~&:&~ q = 0;~ z = 0.3;~ p_z = -0.68,
 \end{eqnarray*}
and investigate their dynamics by computing their  mLE and  SALI. We note that the initial conditions of these orbits  are denoted respectively by red, blue and green dots in Fig.~\ref{energy}(e).

The mLE, $\chi$, is an asymptotic measure characterizing the average rate of growth (or shrinking) of small perturbations to the solutions of a dynamical system and is computed as $\chi = \lim_{t\rightarrow +\infty} \Lambda(t)$, where $\Lambda(t)$ is the so-called finite time mLE
\begin{equation}
    \Lambda(t)= \frac{1}
{t}\ln \left( {\frac{{\left\| {\vec w(t)} \right\|}}
{{\left\| {\vec w(0)} \right\|}}} \right).
\label{eq7}
\end{equation}
In (\ref{eq7}) $\vec w(0)$ and $\vec w(t)$ are the deviation vectors from the studied orbit at times $t=0$ and $t>0$ respectively. It is known that $\chi>0$ denotes chaotic motion, while $\chi=0$ indicates regular orbits \citep{Benettin1,Benettin2,Skokos_lect_lyap}. The value of $\chi$ does not depend on the norm, $\|\, \cdot \, \|$,  used in (\ref{eq7}) and the choice of the initial vector $\vec w(0)$. In our computations we used the common Euclidian norm.

\begin{figure}[h]
\centering
 \begin{center}
\includegraphics[width=0.45\textwidth,height=0.45\textwidth] {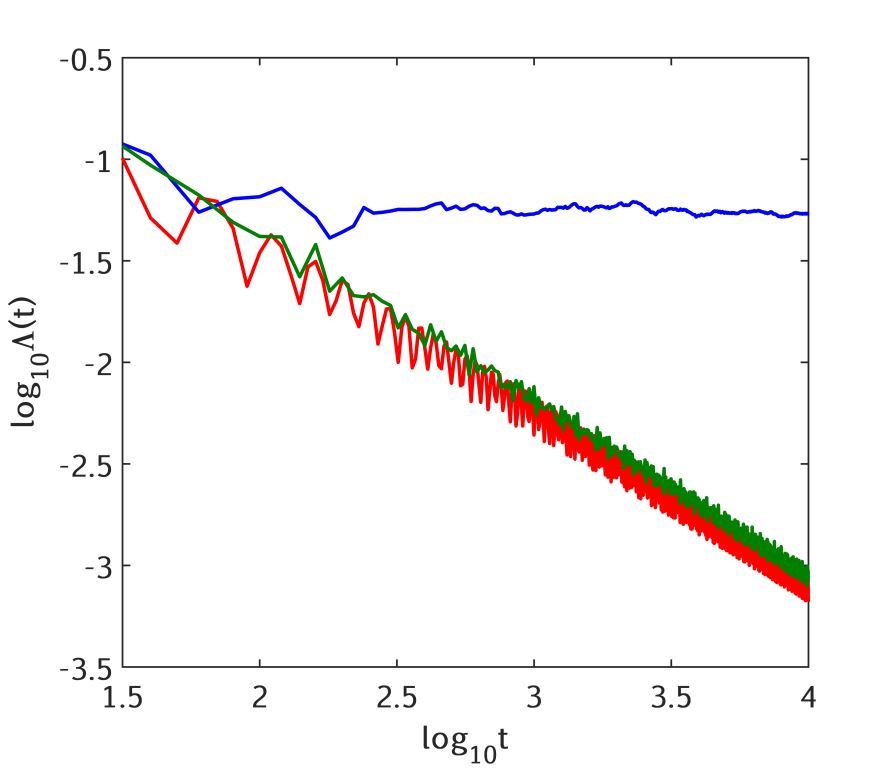}
\end{center}
\caption{Time evolution of the finite time mLE $\Lambda(t)$ in $\log$-$\log$ scale, for the regular orbit A (red curve), the chaotic orbit B (blue curve) and the regular orbit C (green curve), whose initial conditions are shown in Fig.~\ref{energy}(e). }
\label{lyap_conser}
\end{figure}

In Fig.~\ref{lyap_conser}, we plot the time evolution of $\Lambda(t)$ for orbits A (red curve), B (blue curve) and C (green curve). The regular nature of orbits A and C is clearly seen from the results of  Fig.~\ref{lyap_conser} as their finite time mLE tend to zero following a law $\propto t^{-1}$ as is expected for regular motion (see e.g.~Ref.~\citep{Skokos_lect_lyap} and references therein for more details). On the other hand, the evolution of $\Lambda(t)$ for orbit B shows, after some transient phase, a clear tendency to stop decreasing and it seems to stabilize around a positive value of  $\Lambda(t)\approx 10^{-1.3}$. This behavior is indicative of the orbit chaotic nature.

Let us now use the SALI method to characterize the dynamical nature of the orbits A, B and C.
For the computation of the SALI we have to follow the time evolution of two initially different deviation vectors $\vec w_1 (t) = \left( \delta q_1 ,\delta z_1 ,\delta p_{q1} ,\delta p_{z1} \right)$  and $ \vec w_2 (t) = \left( \delta q_2 ,\delta z_2 ,\delta p_{q2} ,\delta p_{z2} \right)$. Then the  SALI at a time $t>0$ is computed as the length of the smallest diagonal  of the parallelogram formed by the  unit vectors $\hat w_1 (t) = w_1 (t) / \left\| {w_1 (t)} \right\|$ and $\hat w_2 (t) = w_2 (t) / \left\| {w_2 (t)} \right\|$ as:
\begin{equation}
\mbox{SALI}(t) = \min \left\{ {\left\| {\hat w_1 (t) + \hat w_2 (t)} \right\|,\left\| {\hat w_1 (t) - \hat w_2 (t)} \right\|} \right\}.
 \label{eq8}
\end{equation}

\begin{figure}[h]
 \centering
  \begin{center}
 \includegraphics[width=0.45\textwidth,height=0.45\textwidth] {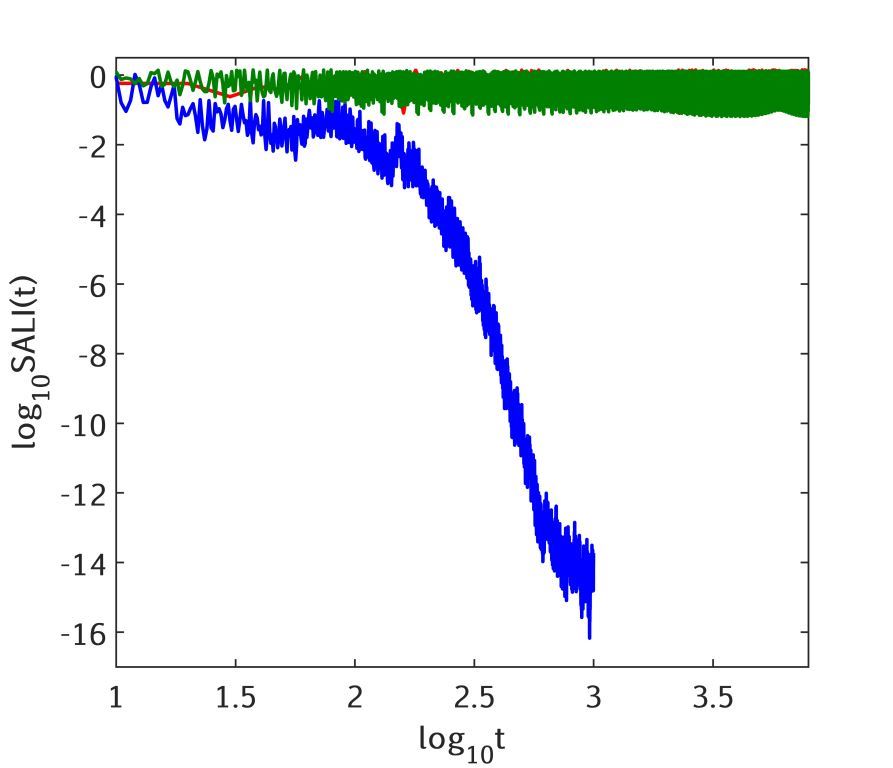}
 \end{center}
 \caption{Time evolution of the SALI$(t)$ in $\log$-$\log$ scale, for the regular orbit A (red curve), the chaotic orbit B (blue curve) and the regular orbit C (green curve), whose initial conditions are shown in Fig.~\ref{energy}(e). We note that the red and green curves practically overlap.} 
 \label{sali_conser}
 \end{figure}

For chaotic orbits, the SALI exhibits a fast decrease to zero (in practice, it reaches quite fast very small values around the computer accuracy, i.e.~$\mbox{SALI} \approx10^{-16}$) because the two deviation vectors tend to become aligned to the direction associated to the mLE, while for regular orbits the index fluctuates around a positive value (for more details see \cite{Skokos_manos} and references therein). These two behaviors are clearly seen in Fig.~\ref{sali_conser} for the regular orbits A (red curve) and C (green curve) and  the chaotic orbit B (blue curve). An important remark here is that the use of the SALI identifies  chaos faster than the computation of the MLE.

\begin{figure}[h]
 \centering
  \begin{center}
 \includegraphics[width=0.45\textwidth,height=0.45\textwidth] {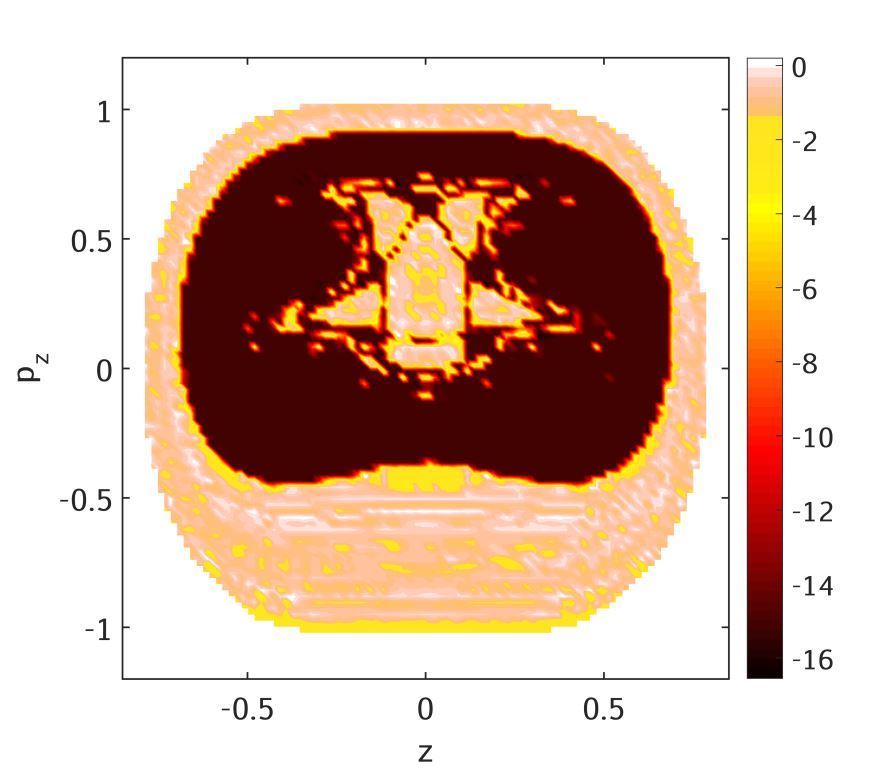}(a)
 \includegraphics[width=0.45\textwidth,height=0.45\textwidth] {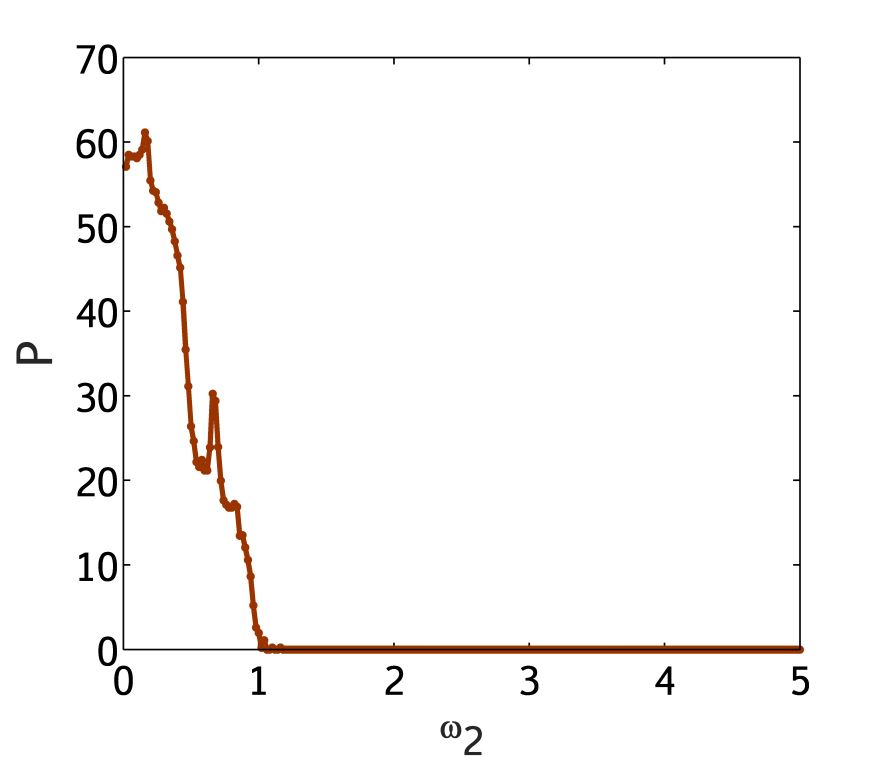}(b)
 \end{center}
 \caption{(a) Regions of different values of the SALI on the PSS defined by $q = 0$, $p_q>0$ of the 2dof Hamiltonian (\ref{eq3}) for $\omega_2 = 0.1$ [Fig.~\ref{energy}(e)]. A grid of  $50 \times 50$ equally spaced initial conditions on the $(y, p_y)$ plane is used. White regions correspond to not-permitted initial conditions. The color scales shown at the right side of panel (a) is used to color each point according to the orbit's $\log_{10}\mbox{SALI}$ value at $t = 3000$.  (b) Percentage $P$ of chaotic orbits (i.e.~orbits having  $\mbox{SALI} \leqslant 10^{ - 8}$ at $t = 3000$) versus the system's mechanical natural frequency $\omega_2$.}
\label{pss_sali}
\end{figure}

A global study of the dynamics of Hamiltonian (\ref{eq3}) can be performed by following the approach implemented in \cite{C35,BCSV_12,BCSPV_12,KKSK_14}. In order to illustrate this approach let us consider a dense grid of initial conditions on the system's PSS for $\omega_2 = 0.1$ [Fig.~\ref{energy}(e)]. Each initial condition is integrated up to $t=3000$ time units and the corresponding point on the PSS is colored according to the value of $\log_{10}\mbox{SALI}$ at the end of the integration. In this way Fig.~\ref{pss_sali}(a) is created, where regions of chaotic behavior corresponding to small values of SALI (colored in black and red), are clearly distinguished from regions with large SALI values where regular motion occurs (colored in pink and yellow). We note that white regions in Fig.~\ref{pss_sali}(a) correspond to not-permitted initial conditions.

Setting a criterion for characterizing an orbit as chaotic the condition $\mbox{SALI} \leqslant 10^{ - 8}$ at the final integration time (which has been used in previous studies \citep{C35,BCSV_12,BCSPV_12,KKSK_14}) we can estimate the percentage $P$ of chaotic orbits for various values of the system's mechanical natural frequency $\omega_2$. The outcome of this analysis is seen in  Fig.~\ref{pss_sali}(b). From the results of this figure we see that for the conservative piezoelectric MEMS the number of chaotic orbits is high, around $60\%$,  for small values of the natural mechanical frequency $\omega_2$ and decreases considerably to zero as  $\omega_2$ becomes large.In particular,  for $\omega_2 = 1$ up to $5$ the system is practically exhibiting only regular motion as no chaotic orbits were found for the resolution of  the used grid of initial conditions.

\section{\label{nonconser_H} Non-conservative piezoelectric MEMS}
In the presence of external forces, which are  explicitly depending on time, the resulting
Hamiltonian system is described by a time dependent Hamiltonian function. In this part of our work we will take into account the effect of not only such external forces but also the influence of  friction or damping phenomena on the dynamics of piezoelectric MEMSs.

\subsection{Hamiltonian function}
\label{sec:sub_Ham_time}
The non-conservative Hamiltonian function is given by (see its derivation in Appendix B):
\begin{eqnarray}
H &=& \left( {\frac{{{\beta _1}}}{{300}}p_q^2 + \frac{{{\gamma _2}{\beta _1}}}{{300{\gamma _1}}}p_z^2} \right){e^{ - \lambda t}} + \left( {\frac{{75}}{{{\beta _1}}}{q^2} + \frac{{75}}{2}{q^4} - \frac{{150{\gamma _1}}}{{{\beta _1}}}qz} \right){e^{\lambda t}}\nonumber\\
 &+& \left( {\frac{{75{\gamma _1}\omega _2^2}}{{{\gamma _2}{\beta _1}}}{z^2} + \frac{{75{\gamma _1}{\beta _2}}}{{2{\gamma _2}{\beta _1}}}{z^4} - \frac{{150}}{{{\beta _1}}}q{E_1}\cos \omega t} \right){e^{\lambda t}},
  \label{eq13}
\end{eqnarray}
where $\lambda$ is a coefficient related to damping, while $E_1$ and $\omega$ denote respectively the external force amplitude and  frequency.

Since the Hamiltonian function (\ref{eq13}) depends  explicitly on time $t$ the system is not any more conservative. The model equations of motions  are:
\begin{eqnarray}
\dot q &=& \frac{{{\beta _1}}}{{150}}{p_q}{e^{ - \lambda t}}  \nonumber \\
\dot z &=& \frac{{{\gamma _2}{\beta _1}}}{{150{\gamma _1}}}{p_z}{e^{ - \lambda t}}  \nonumber \\
{{\dot p}_q} &=& \left( { - \frac{{150}}{{{\beta _1}}}q - 150{q^3} + \frac{{150{\gamma _1}}}{{{\beta _1}}}z + \frac{{150}}{{{\beta _1}}}{E_1}\cos \omega t} \right){e^{\lambda t}}  \nonumber \\
{{\dot p}_z} &=& \left( { - \frac{{150{\gamma _1}\omega _2^2}}{{{\gamma _2}{\beta _1}}}z - \frac{{150{\gamma _1}{\beta _2}}}{{{\gamma _2}{\beta _1}}}{z^3} + \frac{{150{\gamma _1}}}{{{\beta _1}}}q} \right){e^{\lambda t}},
 \label{eq14}
\end{eqnarray}
while the corresponding  variational equations take the form
\begin{eqnarray}
\delta \dot q &=& \frac{{{\beta _1}}}{{150}}\delta {p_q}{e^{ - \lambda t}} \nonumber \\
\delta \dot z &=& \frac{{{\gamma _2}{\beta _1}}}{{150{\gamma _1}}}\delta {p_z}{e^{ - \lambda t}} \nonumber \\
\delta {{\dot p}_q} &=& \left( { - \left( {\frac{{150}}{{{\beta _1}}} + 450{q^2}} \right)\delta q + \frac{{150{\gamma _1}}}{{{\beta _1}}}\delta z} \right){e^{\lambda t}} \nonumber \\
\delta {{\dot p}_z} &=& \left( { - \left( {\frac{{150{\gamma _1}\omega _2^2}}{{{\gamma _2}{\beta _1}}} + \frac{{450{\gamma _1}{\beta _2}}}{{{\gamma _2}{\beta _1}}}{z^2}} \right)\delta z + \frac{{150{\gamma _1}}}{{{\beta _1}}}\delta q} \right){e^{\lambda t}}.
  \label{eq15}
\end{eqnarray}

Let us mention that from (\ref{eq14}) we can easily  obtain the typical set of  equations of a piezoelectric MEMS with damping in the presence of a sinusoidal input voltage [see Eq.~(\ref{bpp44}) in Appendix \ref{Appendix B}]
\begin{eqnarray}
  \ddot q + \lambda \dot q + q + \beta _1 q^3  - \gamma _1 z &=& E_1\cos \omega t \nonumber \\
  \ddot z + \lambda \dot z + \omega _2^2 z + \beta _2 z^3  - \gamma _2 q &=& 0,
  \label{eq16}
\end{eqnarray}

We also note that in the remaining part  of this work Eqs.~(\ref{eq14})-(\ref{eq16}) will be considered with the same parameter values  used in Sect.~\ref{conser_H}. The  values of the additional parameters  are:
\begin{eqnarray*}
\lambda  = 0.05,~~ E_1  = 10.40~~ \textrm{and}~~ \omega  =1~~ \textrm{with}~~ \omega_0  = \omega_e  .
\end{eqnarray*}

\subsection{Effect of the damping coefficient $\lambda$ in the absence of external force}
\label{sec:time_dum}
\begin{figure*}[h]
 \begin{center}
 \includegraphics[width=0.8\textwidth,height=0.6\textwidth] {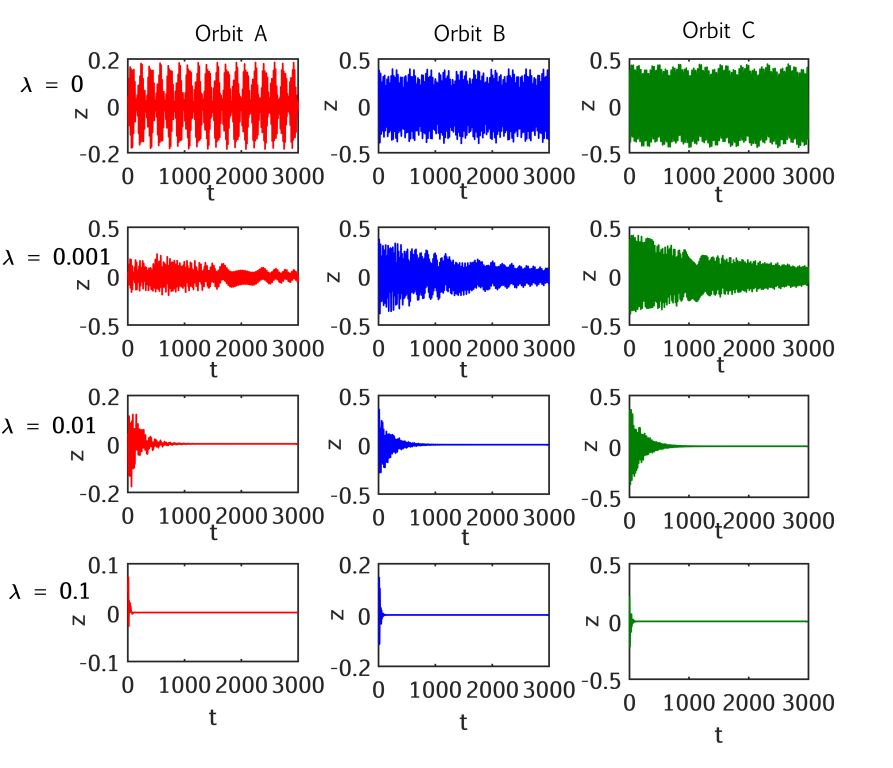}
 \end{center}
 \caption{Time evolution of the $z$ coordinate of orbits A, B and C of Sect.~\ref{sec:level3} for the case of the  non-conservative system  (\ref{eq13}) with $E_1 = 0$ and for various values of the damping parameter $\lambda$.}
\label{TD_timeABC}
\end{figure*}

We start the  investigation of the non-conservative piezoelectric MEMS by studying  the effect of  the damping coefficient on the system's dynamics, assuming that there is no external force acting on it.  For $\lambda=0$ and $E_1 =0$ Hamiltonian  (\ref{eq13}) is equivalent to the conservative system (\ref{eq3}). We fix $E_1 = 0$, slowly vary $\lambda$  from $0$ to nonzero positive values and investigate   the behavior of the three previously studied orbits of the conservative model, namely the regular orbits A and C  and the chaotic orbit B. In Fig.~\ref{TD_timeABC} we present the time evolution of these orbits when the damping coefficient $\lambda$ takes different values. From these results we see that as the damping strength increases, orbits A, B and C practically exhibit the same behavior: all of them  undergo irregular damped oscillations, whose amplitude decreases in time. For higher values of $\lambda$ the dynamics dies out quite fast to the point attractor $q=z=p_q=p_z=0$. Thus, the presence of only damping leads to the eventual death of oscillations in the dissipative piezoelectric MEMS.

\subsection{Effect of the external force amplitude $E_1$ in the absence of damping}
\label{sec:time_force}
\begin{figure}[h]
 \begin{center}
 \includegraphics[width=0.5\textwidth,height=0.65\textwidth] {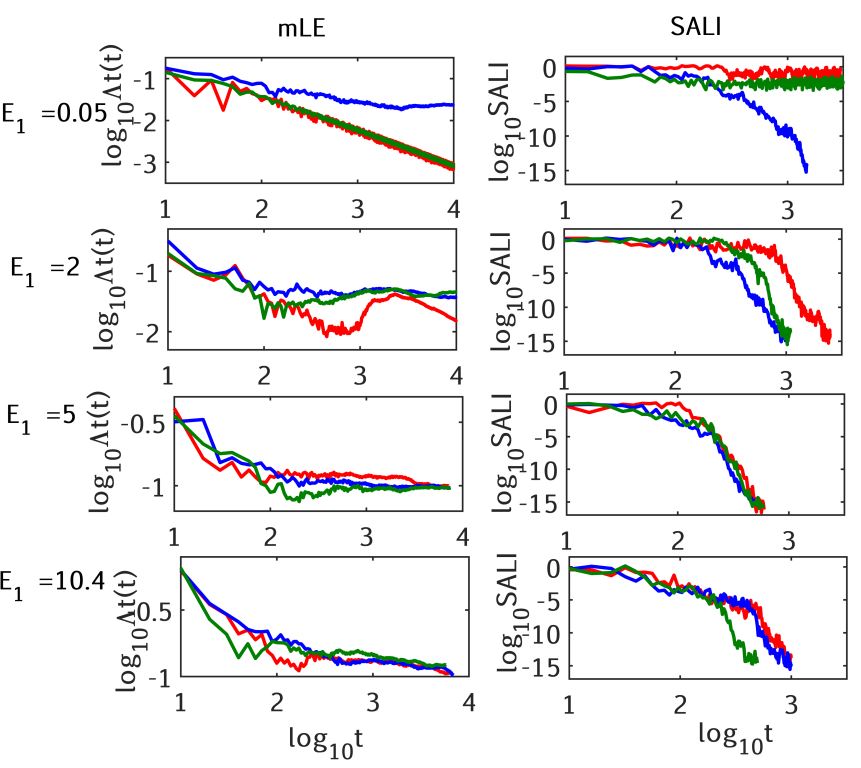}
 \end{center}
 \caption{Time evolution of the finite time  mLE $\Lambda(t)$ and of the SALI$(t)$ in log-log scale for orbits A (red curves),  B (blue curves) and C (green curves) of the  non-conservative system  (\ref{eq13})  for different values of $E_1$, when $\lambda = 0$.}
\label{TD_lyapABC}
\end{figure}

Let us now study the effect of the time periodic external force on the system's dynamics. For this purpose we set $\lambda = 0$ and first investigate the effect of the external force amplitude $E_1$ on the behavior of orbits A, B and C by plotting in Fig.~\ref{TD_lyapABC} the time evolution of their mLE  and SALI for $E_1 =$ 0.05, 2, 5 and 10.4. From these results we see that in all cases the orbits behave chaotically, except from orbits A and C when the amplitude of the external force is very small, i.e.~$E_1=0.05$. In these particular cases, $\Lambda(t)$ tends to zero showing a continuous decrease to smaller, positive  values, while the SALI fluctuates around a non-zero positive value. Both these behaviors indicate the regular nature of these orbits. In all the other cases of Fig.~\ref{TD_lyapABC} $\Lambda(t)$ saturates to positive values, which increase as $E_1$ becomes larger. In agreement to this behavior the SALI of all chaotic orbits decreases very fast to zero. We see that SALI reaches very small values (e.~g.~$\mbox{SALI} = 10^{-8}$) faster when $E_1$ increases. The behavior of both the finite mLE and the SALI in Fig.~\ref{TD_lyapABC} clearly indicates that the increase of the external force amplitude makes the system more chaotic.

\begin{figure}[h]
 \centering
  \begin{center}
 \includegraphics[width=0.5\textwidth,height=0.5\textwidth] {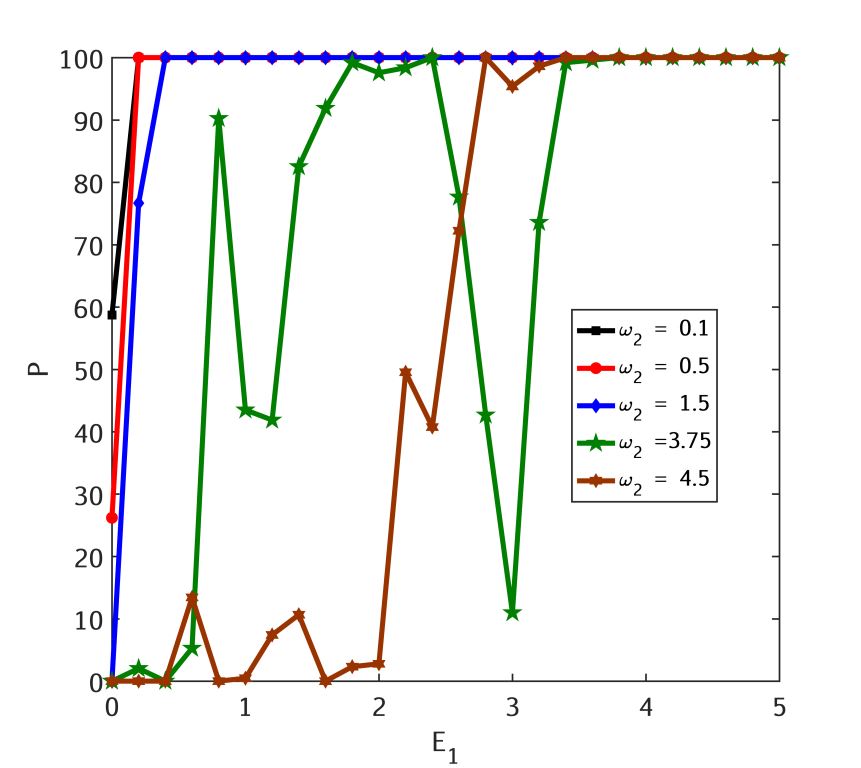}
 \end{center}
 \caption{Percentage $P$ of chaotic orbits (i.e.~orbits having  $\mbox{SALI} \leqslant 10^{ - 8}$ at $t = 3000$) versus the external force amplitude $E_1$ for four ensembles of orbits having  $\omega_2=0.1$, 0.5, 1.5, 3.75  and 4.5 for $\lambda=0$.  }
\label{TD_percentage}
\end{figure}

Since the decrease of the SALI to small values, like $\mbox{SALI} = 10^{-8}$, is sufficient to characterize an orbit as chaotic, we use this criterion to perform a more general investigation of the dynamics of the piezoelectric MEMS in the absence of damping by following the evolution of the percentage $P$ of chaotic orbits as a function of $E_1$ for some particular cases. In other words, we perform a similar analysis to the one presented in Fig.~\ref{pss_sali}(b). In particular, we integrate up to $t=3\,000$ the initial conditions used in  Fig.~\ref{pss_sali}(b) for $\omega_2=0.1$, 0.5, 1.5, 3.75  and 4.5 [see Fig.~\ref{energy}(e)],  considering the non-conservative system (\ref{eq13}), and find out how $P$ depends on $E_1$ (Fig.~\ref{TD_percentage}).

From the results of Fig.~\ref{TD_percentage} we see that when $E_1=0$ the percentage $P$ of chaotic orbits is large for $\omega_2=0.1$ and $0.5$ ($P \approx 60\%$ and $P \approx 26\%$ respectively) and equal to zero for $\omega_2=1.5$, $3.75$ and $4.5$ [Fig.~\ref{pss_sali}(b)]. As $E_1$ increases, the percentage $P$ of chaotic orbits for the cases $\omega_2=0.1$, $0.5$ and $1.5$ grows fastly and saturates to $P = 100\%$ for $E_1 \gtrsim 0.2$ while for other cases ($\omega_2 = 3.75$ and $4.5$), we observe some variations of $P$ before it reaches the saturation from $E_1 \gtrsim 3.4$.  These results clearly indicate that chaos eventually dominates the dynamics of the piezoelectric MEMS when the external force's amplitude grows.

\subsection{Effect of the coexistence of damping and external force}
\label{sec:time_dam_force}
In order to investigate the dynamics of the piezoelectric MEMS in the case of the simultaneous presence of damping and of an external driving force, we solve numerically Eqs.~(\ref{eq16}) instead of Eqs.~(\ref{eq14}) due to numerical instabilities in the solution of the latter system because quantities $p_q$ and $p_z$ become progressively very large as $\lambda$ increases. In particular, considering the new variables $u=\dot{q}$ and $v=\dot{z}$ Eqs.~(\ref{eq16})
can be rewritten as:
\begin{eqnarray}
\dot q &=& u \nonumber\\
\dot u &=&  - \lambda u - q - {\beta _1}{q^3} + {\gamma _1}z + {E_1}\cos \omega t \nonumber\\
\dot z &=& v \\
\dot v &=&  - \lambda v - \omega _2^2z - {\beta _2}{z^3} + {\gamma _2}q, \nonumber
\label{eq18}
\end{eqnarray}
while the corresponding  variational equations take the form
\begin{eqnarray}
\dot{\delta  q} &=& \delta u\nonumber\\
\dot{\delta u} &=&  - \lambda \delta u - \left( {1 + 3{\beta _1}{q^2}} \right)\delta q + {\gamma _1}\delta z \nonumber\\
\dot{\delta z} &=& \delta v\\
\dot{ \delta  v} &=&  - \lambda \delta v - \left( {\omega _2^2 + 3{\beta _2}{z^2}} \right)\delta z + {\gamma _2}\delta q. \nonumber
\label{eq19}
\end{eqnarray}

\begin{figure}[h]
\centering
 \begin{center}
\includegraphics[width=0.45\textwidth,height=0.45\textwidth] {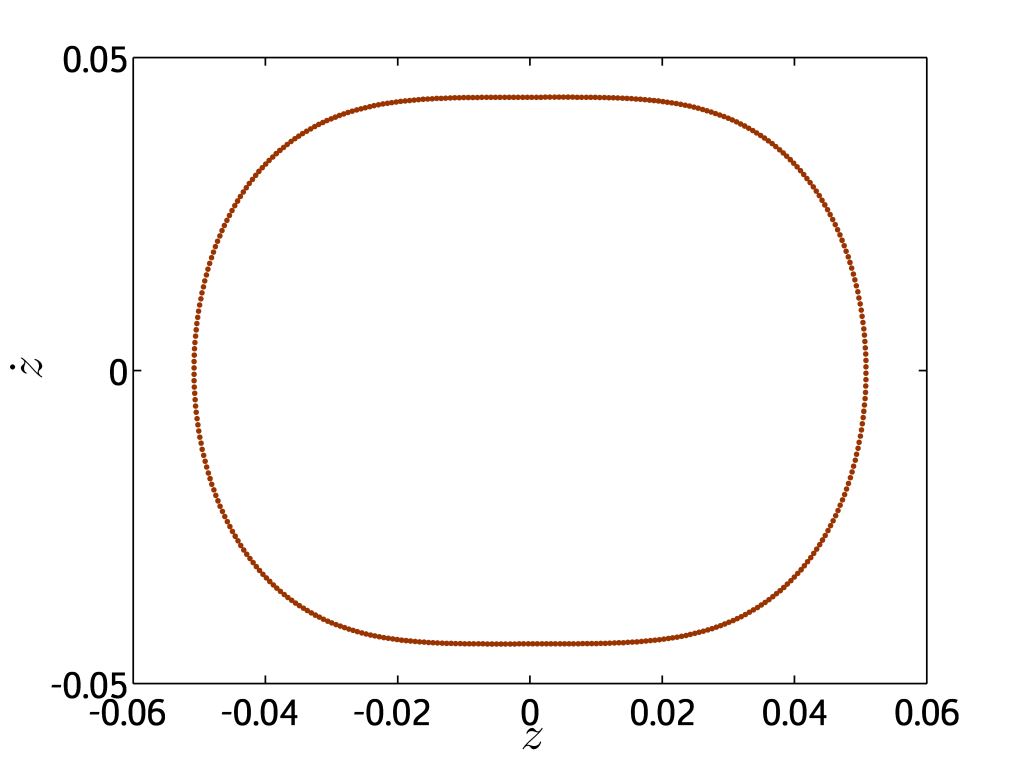}
\end{center}
\caption{Projection of PSS of the non-conservative piezoelectric MEMS  (\ref{eq13}) on the $(z, \dot z)$ plane for  $E_1 = 0.05$  and $\lambda = 0.05$. }
\label{PSS}	
\end{figure}

As we saw in Sect.~\ref{sec:time_dum} damping results to the eventual disappearance of dynamical evolution, while the external forcing of the system (Sect.~\ref{sec:time_force}) leads to extensive chaos. In order to check if we can obtain different dynamical behaviors when both factors (damping and external force) are present, we initially consider the case where the damping coefficient is $\lambda=0.05$ and the value of the external force amplitude is $E_1 = 0.05$. To analyze the system's dynamics for these parameter values we consider its PSS. We remind that the PSS in a periodically driven system like (\ref{eq13}) is obtained by registering the orbital coordinates (in our case the vector $\left( q,z,\dot{q}, \dot{z}\right)$ ) in a stroboscopic manner, i.e.~at each period $T$ of the external force, or in other words at times $t=i T= i \frac{2 \pi}{\omega}$, $i=1,2,\ldots$.
In Fig.~\ref{PSS} we see the projection of the PSS on the $(z, \dot{z})$ plane for orbits with initial conditions $q=0$, $\dot q =4.138$, while $z$ and $\dot{z}$ are given on a grid of $50 \times 50$ equally spaced points in the region $-0.06 \leq z < 0.06$, $-0.05 \leq \dot{z} < 0.05$. In order to discard the initial, transient phase of the dynamics we plot in Fig.~\ref{PSS} only the points of the considered orbits for $1\,500 \leq t \leq 3\, 500$.

\begin{figure}[h]
 \centering
  \begin{center}
 \includegraphics[width=0.45\textwidth,height=0.45\textwidth] {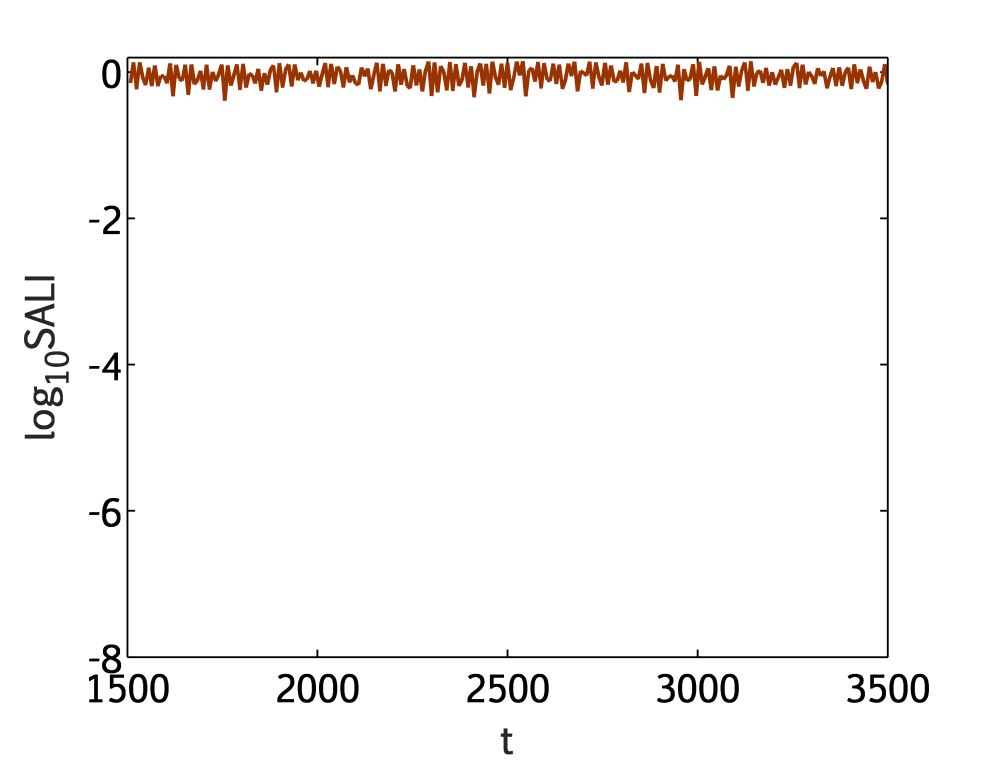}
 \end{center}
 \caption{The evolution of  the $\mbox{SALI}(t)$,  for $1\,500 \leq t \leq 3\, 500$, of the orbit with initial conditions  $q=0$, $z=0.04935$, $\dot{q}=4.138$, $\dot{z}=0.01354$ for the non-conservative piezoelectric MEMS  (\ref{eq13}) with $E_1 = 0.05$  and $\lambda = 0.05$. }
\label{mLE_sali_E50over6}
\end{figure}

From the results of Fig.~\ref{PSS} we notice that the consequents of all considered initial conditions are distributed on a smooth curve, indicating the existence of an attractor on which regular motion takes place. Thus, the particular interplay of low damping and small amplitude of the external driving force leads the piezoelectric MEMS to regular behavior. This behavior is also evident by the time evolution of  the SALI in Fig.~\ref{mLE_sali_E50over6} of one particular orbit considered in Fig.~\ref{PSS}, namely the one with initial conditions $q=0$, $z=0.04935$, $\dot{q}=4.138$, $\dot{z}=0.01354$. Again, excluding from our analysis an initial transient phase, we present in Fig.~\ref{mLE_sali_E50over6} the evolution of  $\mbox{SALI}(t)$ for $t\geq 1\,500$.

\begin{figure}[ht]
\centering
 \begin{center}
\includegraphics[width=0.5\textwidth,height=0.5\textwidth] {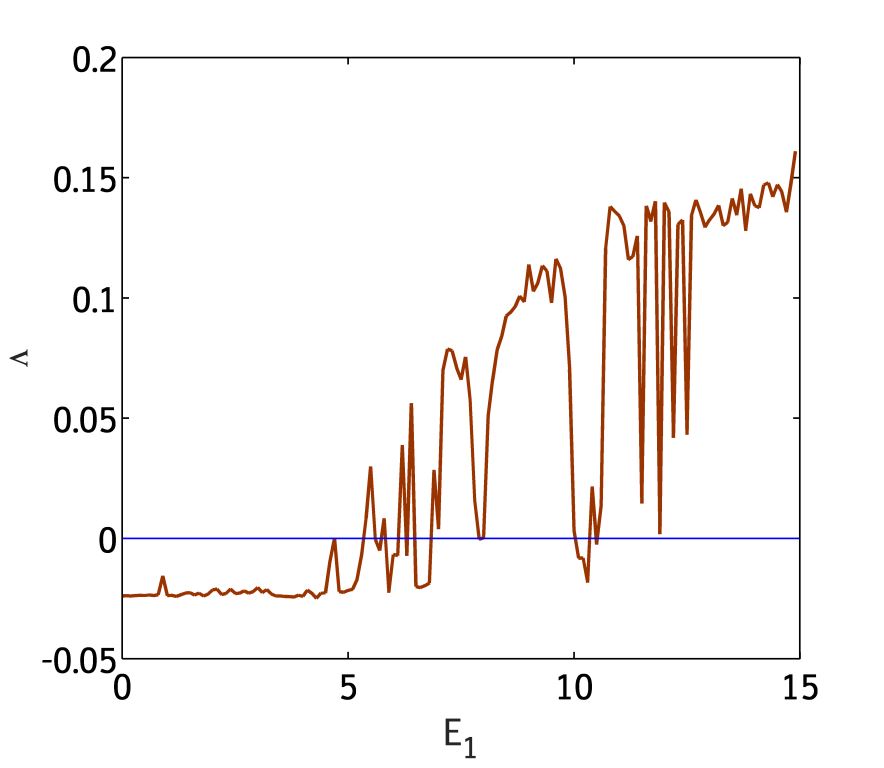}
\end{center}
\caption{The value of the finite time mLE $\Lambda$ at $t=3\, 500$, of orbits with initial conditions $q = 0$, $z =-0.042$, $\dot{q}=4.138$, $\dot{z}=0.31$, as a function of the parameter $E_1$  of the time dependent system (\ref{eq13}) with $\lambda=0.05$. The line $\Lambda=0$ is also plotted.}
\label{lyap_final}
\end{figure}

In order to investigate the robustness of the appearance of regular motion when both damping and external force are present, we keep $\lambda = 0.05$ and examine how the external force's amplitude affects the behavior of one representative initial condition. In particular, we consider the initial condition of orbit A ($q = 0$, $z =-0.042$, $\dot{q}=4.138$, $\dot{z}=0.31$), which corresponds to a regular orbit of the conservative system (\ref{eq3}) [see Fig.~\ref{energy}(e)], and register the value of the finite time mLE $\Lambda$ at $t=3\, 500$ for $0\leq E_1 \leq 15$. The outcome of this process is seen in Fig.~\ref{lyap_final}. From the results of this figure we understand that the considered initial condition leads to a more complex dynamics. Indeed  for $0\leq E_1 \lesssim 5$ a regular motion appears as the corresponding finite time mLE is negative, followed by a series of transitions between regular and chaotic motion up to the value $E_1 \approx 10.3$. For rather large values of the  external force amplitude $E_1$ only chaotic dynamics takes place. We can also remark that for the studied case the coexistence of damping and external force leads  the MEMS to be more chaotic than regular.

\begin{figure}[ht]
\centering
 \begin{center}
\includegraphics[width=0.5\textwidth,height=0.5\textwidth] {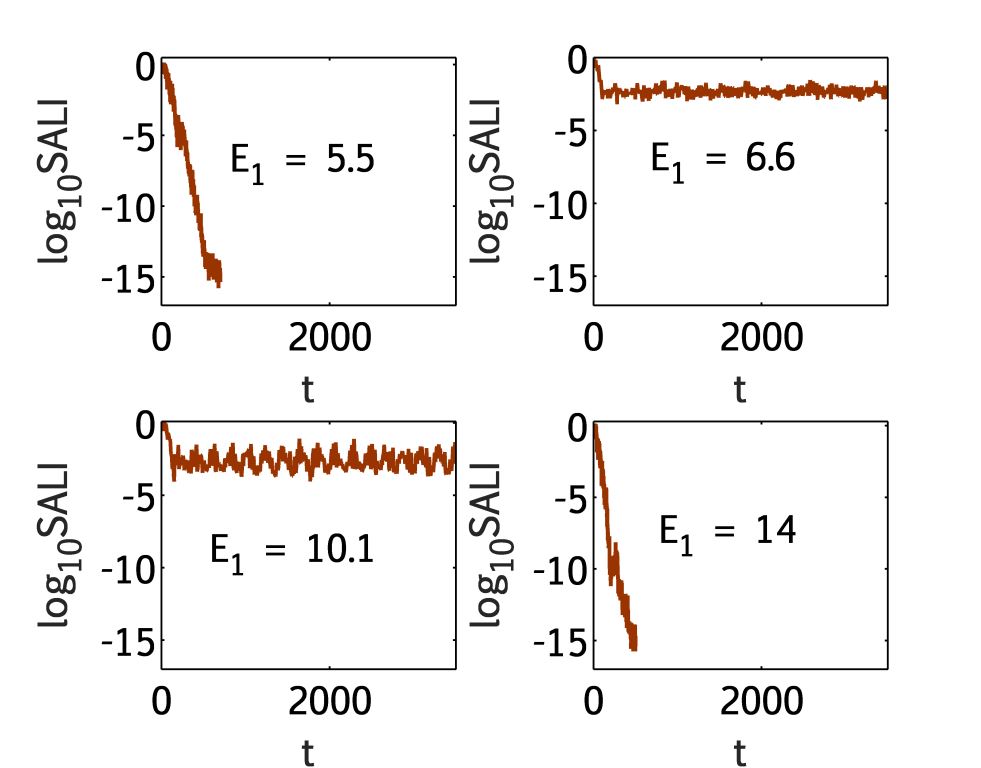}
\end{center}
\caption{Time evolution of the SALI$(t)$ for some particular cases of Fig.~\ref{lyap_final}: $E_1=6.6$, $10.1$ (regular motion),  and $E_1=5.5$, $14$ (chaotic motion).}
\label{TD_sali4E1}
\end{figure}

The results of Fig.~\ref{lyap_final} are also in agreement with the computations of the SALI method. This is seen in Fig.~\ref{TD_sali4E1} where the time evolution of the SALI for some particular cases of Fig.~\ref{lyap_final} are seen, namely the ones for $E_1=5.5$, 6.6, 10.1 and 14. The two cases $E_1 = 6.6 $ and $10.1$ correspond to regular motion as the SALI remains almost constant and positive (the corresponding finite time mLEs in  Fig.~\ref{lyap_final} are negative), while the two other cases ($E_1=5.5$ and 14) correspond to chaotic orbits because their SALI decreases to zero ($\mbox{SALI} \lesssim 10^{-15}$) very fast (for these cases $\Lambda>0$ in   Fig.~\ref{lyap_final}). We see again here that chaotic motion is very fast identified by the SALI method.

\newpage
\section{\label{conclu}Summary and conclusions}
We numerically investigated the dynamics of a piezoelectric MEMS when the system is considered isolated from its environment, and consequently it is described by a conservative Hamiltonian model, as well as when damping and/or external forces are taken into account, leading to a time dependent Hamiltonian system. In our work we studied the behavior of individual orbits of these systems using the Poincar\'{e} Surface of Section technique to visualize their dynamics and evaluated appropriate chaos indicators, namely the  maximum Lyapunov Exponent and the Smaller Alignment Index, to quantify their chaoticity. In addition, performing extensive simulations of ensembles of many orbits we studied the global dynamics of the considered models.

Our results show that in the case of the conservative piezoelectric MEMS (\ref{eq3}) the system predominantly exhibits regular motion for large values of the natural frequency of the system's mechanical part i.e.~$\omega_2 > 1$. However for small values of $\omega_2$ a significant percentage of the considered initial conditions lead to chaotic motion. This percentage diminishes with the increase of $\omega_2$.

On the other hand, the non-conservative version of the piezoelectric MEMS (\ref{eq13}) exhibited a reacher dynamical behavior, which is mainly influenced by two factors: the energy loss due to damping and friction, quantified by the damping coefficient $\lambda$ in (\ref{eq13}), and the energy gain through the action of an external, time periodic force of variable amplitude $E_1$. In all studied cases the motion was eventually led on regular or chaotic attractors. More specifically, when only damping was present the motion was dying out and the system ended up to a point attractor of zero energy, while the presence of the external force in the absence of damping resulted to extended chaos for sufficiently large values of $E_1$. The coexistence of both energy loss and gain led to more complicated behaviors with the system undergoing transitions from regular to chaotic dynamics.

Our results shed some light on the complicated dynamical behavior the piezoelectric MEMS can exhibit, indicating the necessity of a more detailed study of the system's dynamics in the parameter space $(\lambda, E_1)$, a task we intend to undertake in a future publication. The current work, along with future studies of realistic piezoelectric MEMSs in the same vein, will be helpful for understanding the functioning of devices using piezoelectric actuators and eventually improving their efficiency; a goal which is of significant practical importance.

As a final remark let us note that the SALI method has been mainly used to date for studying conservative systems, although some applications of the index to time dependent models have already appeared in the literature \citep{Huang_wu,Huang_zhou,Huang_cao,Manos_al_jpa,Huang_wu_lect,HSP19}. Our study adds value to these, rather few works, as it provides additional, clear evidences that the SALI is an easy to compute, reliable and very efficient chaos detection technique also for time dependent systems.


\nonumsection{Acknowledgments} \noindent M.~V.~Tchakui wishes to acknowledge the support of the African-German Network of Excellence in Science (AGNES) for receiving a Mobility Grant in 2016, which allowed her to visit the Department of Mathematics and Applied Mathematics (MAM) of the University of Cape Town; the Grant is generously sponsored by German Federal Ministry of Education and Research and supported by the Alexander von Humboldt Foundation. She also thanks MAM for its hospitality during her visit. Ch.~Skokos thanks the Max Planck Institute for the
Physics of Complex Systems in Dresden, Germany for its hospitality during his visit there in the second half of 2018, when part of this work was carried out.


\appendix{Derivation of the conservative Hamiltonian function}
\label{Appendix A}
The Lagrangian of Eq.~(\ref{eq1}) can be rewritten as:
\begin{eqnarray}
\Gamma\left( {\dot q(\tau ),\dot z(\tau ),q(\tau ),z(\tau )} \right) &=& \alpha {\dot z^2} + {b_1}{\dot q^2} - c{z^2} - d{z^4} - e{q^2} \nonumber\\  &-& f{q^4} + gqz,
 \label{app1}
\end{eqnarray}
where

\begin{eqnarray}
\alpha  &=& \frac{1}{2}M,~ {b_1} = \frac{1}{2}L,~ c = \frac{1}{2}\left( {{K_0} + \frac{{{K_a}{b^2}}}{{1 - {k^2}}}} \right),~d = \frac{1}{4}{K_1}\label{appa}
\nonumber\\
e &=& \frac{1}{2}\left( {\frac{1}{{{C_{0l}}}} + \frac{1}{{C(1 - {k^2})}}} \right),~f = \frac{1}{4}{\beta _{e0}},~g = \frac{{n{d_{33}}{K_a}b}}{{C(1 - {k^2})}}. \label{appb}
 \label{app10}
\end{eqnarray}
By considering the notation $q_1=q$, $q_2=z$, we obtain the Hamiltonian function corresponding to Eq.~(\ref{app1}) through the relation
\begin{eqnarray*}
H = \sum\limits_{i = 1}^2 {{{\dot q}_i}{p_i} - \Gamma} ~~\textrm{with}~~{p_i} = \frac{{\partial \Gamma}}{{\partial {{\dot q}_i}}},\label{appc}
\end{eqnarray*}
to be:
\begin{eqnarray}
H\left( {\dot q(\tau ),\dot z(\tau ),q(\tau ),z(\tau )} \right) &=& \alpha {\dot z^2} + {b_1}{\dot q^2} + c{z^2} + d{z^4} + e{q^2}\nonumber\\ &+& f{q^4} - gqz.
 \label{app2}
\end{eqnarray}
The Lagrange equations of motion in the conservative case are obtained from the relation:

\begin{eqnarray*}
\frac{d}{{d\tau }}\left( {\frac{{\partial \Gamma}}{{\partial {{\dot q}_i}}}} \right) - \frac{{\partial \Gamma}}{{\partial {q_i}}} = 0, \,\,\,\,\,\, i=1,2,
\label{app}
\end{eqnarray*}

are
\begin{eqnarray}
\frac{{{d^2}q}}{{d{\tau ^2}}} + \frac{e}{{{b_1}}}q + \frac{{2f}}{{{b_1}}}{q^3} - \frac{g}{{2{b_1}}}z = 0\nonumber\\
\frac{{{d^2}z}}{{d{\tau ^2}}} + \frac{c}{\alpha }z + \frac{{2d}}{\alpha }{z^3} - \frac{g}{{2\alpha }}q = 0.
\label{app3}
\end{eqnarray}
Let us consider a dimensional time $t$  so that:
\begin{eqnarray}
t=\tau \omega _e ~~\textrm{with}~~\omega _e^2 = \frac{e}{{{b_1}}}.
\label{appt}
\end{eqnarray}
 The system (\ref{app3}) can be rewritten in a dimensionless form as:
\begin{eqnarray}
\frac{{{d^2}q}}{{d{t^2}}} + q + \frac{{2f}}{{\omega _e^2{b_1}}}{q^3} - \frac{g}{{2\omega _e^2{b_1}}}z = 0\nonumber\\
\frac{{{d^2}z}}{{d{t^2}}} + \frac{c}{{\omega _e^2\alpha }}z + \frac{{2d}}{{\omega _e^2\alpha }}{z^3} - \frac{g}{{2\omega _e^2\alpha }}q = 0,
\label{app4}
\end{eqnarray}
and with new parameters as:
\begin{eqnarray}
\ddot q + q + {\beta _1}{q^3} - {\gamma _1}z = 0\nonumber\\
\ddot z + \omega _2^2z + {\beta _2}{z^3} - {\gamma _2}q = 0,
\label{app5}
\end{eqnarray}
where
\begin{eqnarray}
{\beta _1} = \frac{{2f}}{{\omega _e^2{b_1}}},~{\gamma _1} = \frac{g}{{2\omega _e^2{b_1}}},~\omega _2^2 = \frac{c}{{\omega _e^2\alpha }}\nonumber\\
{\beta _2} = \frac{{2d}}{{\omega _e^2\alpha }},~{\gamma _2} = \frac{g}{{2\omega _e^2\alpha }}.
\label{app44}
\end{eqnarray}
\newline
In these new dimensionless variables the Lagrangian and Hamiltonian functions respectively become:
\begin{eqnarray*}
\Gamma\left( {\dot q(t ),\dot z(t ),q(t ),z(t )} \right) &=& \alpha \omega _e^2{\dot z^2} + {b_1}\omega _e^2{\dot q^2} - c{z^2} - d{z^4} - e{q^2}\nonumber\\ &-& f{q^4} + gqz,\nonumber\\
H\left( {\dot q(t ),\dot z(t ),q(t ),z(t )} \right) &=& \alpha \omega _e^2{\dot z^2} + {b_1}\omega _e^2{\dot q^2} + c{z^2} + d{z^4} + e{q^2}\nonumber\\ &+& f{q^4} - gqz.
\end{eqnarray*}
We have:
\begin{eqnarray*}
{p_q} = \frac{{\partial \Gamma}}{{\partial \dot q}} = 2\omega _e^2{b_1}\dot q \to \dot q = \frac{{{p_q}}}{{2\omega _e^2{b_1}}}\nonumber\\
{p_z} = \frac{{\partial \Gamma}}{{\partial \dot z}} = 2\omega _e^2\alpha \dot z \to \dot z = \frac{{{p_z}}}{{2\omega _e^2\alpha }},
\end{eqnarray*}
 so the dimensionless Hamiltonian function can be rewritten using the momentum variables as:
\begin{eqnarray}
H\left( {p_q(t),p_z(t),q(t),z(t)} \right) &=& \frac{{p_z^2}}{{4\alpha \omega _e^2}} + \frac{{p_q^2}}{{4{b_1}\omega _e^2}} + c{z^2} + d{z^4}\nonumber\\ &+& e{q^2} + f{q^4} - gqz.
 \label{app7}
\end{eqnarray}
In order to express the parameters $\alpha$, $b_1$, $c$, $d$, $e$, $f$ and $g$ as function of new parameters $\beta _1$, $\gamma _1$, $\omega _2$, $\beta_2$ and $\gamma _2$, we solve Eq.~(\ref{app44}) by considering the relations ${b_1} = \frac{1}{2}L$ and $f = \frac{1}{4}{\beta _{e0}}$ from Eqs.~(\ref{app10}) and the values $L=1, \beta_{e0}=150$ from Table \ref{tab:value}. We thus get:
\begin{eqnarray}
\alpha  = \frac{{{\gamma _1}}}{{2{\gamma _2}}},~~{b_1} = \frac{1}{2},~~c = \frac{{75{\gamma _1}\omega _2^2}}{{{\gamma _2}{\beta _1}}},~~d = \frac{{75{\gamma _1}{\beta _2}}}{{2{\gamma _2}{\beta _1}}}\nonumber\\
e = \frac{{75}}{{{\beta _1}}},~~f = \frac{{75}}{2},~~g = \frac{{150{\gamma _1}}}{{{\beta _1}}},~~\omega _e^2 = \frac{{150}}{{{\beta _1}}}.
\label{app8}
\end{eqnarray}
Substituting Eqs.~(\ref{app8}) in Eq.~(\ref{app7}) we get the conservative Hamiltonian function as seen in Eq.~(\ref{eq3}):
\begin{eqnarray}
H\left( {p_q,p_z,q,z} \right) &=& \frac{{{\beta _1}}}{{300}}p_q^2 + \frac{{{\gamma _2}{\beta _1}}}{{300{\gamma _1}}}p_z^2 + \frac{{75}}{{{\beta _1}}}{q^2} + \frac{{75}}{2}{q^4} + \frac{{75{\gamma _1}\omega _2^2}}{{{\gamma _2}{\beta _1}}}{z^2}\nonumber\\
 &+& \frac{{75{\gamma _1}{\beta _2}}}{{2{\gamma _2}{\beta _1}}}{z^4} - \frac{{150{\gamma _1}}}{{{\beta _1}}}qz.
\label{app9}
\end{eqnarray}

\appendix{\label{Appendix B}Derivation of the conservative Hamiltonian function}
The virtual work of the non-conservative forces is given by:
\begin{eqnarray}
\delta {W_{nc}} = E(\tau )\delta q,
 \label{bpp1}
\end{eqnarray}
where $E(\tau)$ is the voltage source intensity given in Eq.~(\ref{voltage}).
The Lagrange equations in the non-conservative case obtained from relation
\begin{eqnarray}
\frac{d}{{d\tau }}\left( {\frac{{\partial \Gamma }}{{\partial {{\dot q}_i}}}} \right) - \frac{{\partial \Gamma }}{{\partial {q_i}}} + \frac{{\partial \Lambda }}{{\partial {{\dot q}_i}}} = {F_{qi}},
 \label{bpp2}
\end{eqnarray}
where $F_{qi}$ denotes the non-conservative forces acting on the system (namely $E(\tau)$  in this particular case), while $\Lambda$ is the total dissipation function expressed in Eq.~(\ref{dissipation})  and $\Gamma$ is the Lagrangian ofEq.~(\ref{app1}), can be written as:
\begin{eqnarray}
\frac{{{d^2}q}}{{d{\tau ^2}}} + \frac{R}{{2{b_1}}}\frac{{dq}}{{d\tau }} + \frac{e}{{{b_1}}}q + \frac{{2f}}{{{b_1}}}{q^3} - \frac{g}{{2{b_1}}}z &=& {E_e}\cos {\omega _0}\tau \nonumber\\
\frac{{{d^2}z}}{{d{\tau ^2}}} + \frac{{{\lambda _{m0}}}}{{2\alpha }}\frac{{dz}}{{d\tau }} + \frac{c}{\alpha }z + \frac{{2d}}{\alpha }{z^3} - \frac{g}{{2\alpha }}q &=& 0.
 \label{bpp3}
\end{eqnarray}

In dimensionless form using the transformation of Eq.~(\ref{appt}), Eq.~(\ref{bpp3}) gives:
\begin{eqnarray}
\frac{{{d^2}q}}{{d{t^2}}} + \frac{R}{{2{b_1}{\omega _e}}}\frac{{dq}}{{d\tau }} + q + \frac{{2f}}{{{b_1}\omega _e^2}}{q^3} - \frac{g}{{2{b_1}\omega _e^2}}z &=& \frac{{{E_e}}}{{\omega _e^2}}\cos \frac{{{\omega _0}}}{{{\omega _e}}}t \nonumber\\
\frac{{{d^2}z}}{{d{\tau ^2}}} + \frac{{{\lambda _{m0}}}}{{2\alpha {\omega _e}}}\frac{{dz}}{{d\tau }} + \frac{c}{{\alpha \omega _e^2}}z + \frac{{2d}}{{\alpha \omega _e^2}}{z^3} - \frac{g}{{2\alpha \omega _e^2}}q &=& 0.
 \label{bpp4}
\end{eqnarray}
For simplicity reason, we choose $\lambda _{m0}$ so that
\begin{eqnarray*}
 \frac{R}{{2{b_1}{\omega _e}}} = \frac{{{\lambda _{m0}}}}{{2\alpha {\omega _e}}}= \lambda.
 \label{bpp44}
\end{eqnarray*}
Thus, parameter $\lambda$ denotes the dissipation coefficient in the piezosystem. Then Eqs.~(\ref{bpp4}) can be rewritten as:
\begin{eqnarray}
\ddot q + {\lambda }\dot q + q + {\beta _1}{q^3} - {\gamma _1}z &=& {E_1}\cos \omega t \nonumber\\
 \ddot z + {\lambda}\dot z + \omega _2^2z + {\beta _2}{z^3} - {\gamma _2}q &=& 0,
 \label{bpp5}
\end{eqnarray}
where
\begin{eqnarray}
{E_1} = \frac{{{E_e}}}{{\omega _e^2}}~~\textrm{and}~\omega  = \frac{{{\omega _0}}}{{{\omega _e}}}~~\textrm{with}~{E_e}=\frac{E_{e0}}{L}.
 \label{bpp6}
\end{eqnarray}
Following the analysis of e.g.~\cite{Goldstein} on the  Lagrangian and Hamiltonian description of  dissipative systems  the Lagrangian function of the piezoelectric system and the corresponding Hamiltonian, when we take into account friction and external time periodic forces, can be written  in dimensionless variables as:
\begin{eqnarray}
\Gamma(t) &=& ( \alpha \omega _e^2{{\dot z}^2} + {b_1}\omega _e^2{{\dot q}^2} - c{z^2} - d{z^4} - e{q^2} - f{q^4} + gqz){e^{\lambda t}}\nonumber\\ &+& (q\omega _e^2{E_1}\cos \omega t){e^{\lambda t}},
 \label{bpp7}
\end{eqnarray}
\begin{eqnarray}
H(t) &=& ( \alpha \omega _e^2{{\dot z}^2} + {b_1}\omega _e^2{{\dot q}^2} + c{z^2} + d{z^4} + e{q^2} + f{q^4} - gqz){e^{\lambda t}}\nonumber\\ &-& (q\omega _e^2{E_1}\cos \omega t) {e^{\lambda t}}.
 \label{bpp8}
\end{eqnarray}
Using momentum coordinates $p_q$ and $p_z$, and replacing the parameters $\alpha$, $b_1$, $c$, $d$, $e$, $f$ and $g$ through the expression provided in Eq.~(\ref{app8})  the non-conservative Hamiltonian (\ref{bpp8}) becomes:
\begin{eqnarray}
H(t) &=& \left( {\frac{{{\beta _1}}}{{300}}p_q^2 + \frac{{{\gamma _2}{\beta _1}}}{{300{\gamma _1}}}p_z^2} \right){e^{ - \lambda t}} + \left( {\frac{{75}}{{{\beta _1}}}{q^2} + \frac{{75}}{2}{q^4} + \frac{{75{\gamma _1}\omega _2^2}}{{{\gamma _2}{\beta _1}}}{z^2}} \right){e^{\lambda t}}\nonumber\\
 &-& \left( { - \frac{{75{\gamma _1}{\beta _2}}}{{2{\gamma _2}{\beta _1}}}{z^4} + \frac{{150{\gamma _1}}}{{{\beta _1}}}qz + \frac{{150}}{{{\beta _1}}}q{E_1}\cos \omega t} \right){e^{\lambda t}}.
 \label{bpp9}
\end{eqnarray}

\bibliographystyle{ws-ijbc}
\bibliography{references_n}
\end{document}